\documentclass[a4paper,fleqn,usenatbib]{mnras}

\usepackage[T1]{fontenc}
\usepackage{ae,aecompl}
\usepackage{graphicx}	
\usepackage{amsmath}	
\usepackage{amssymb}	
\usepackage{cleveref}   
\usepackage{hyperref}   

\title[Dynamical status of NGC 6656]{The radial distribution of Blue stragglers in Galactic Globular Cluster NGC 6656 $-$ Clues on the dynamical status}

\author[Gaurav Singh and R. K. S. Yadav]{Gaurav Singh \thanks{E-mail:
gaurav@aries.res.in} and R. K. S. Yadav \thanks{E-mail:
rkant@aries.res.in}\\ Aryabhatta Research Institute of Observational Sciences, Manora Peak, Nainital 263002, India\\ Department of Physics and Astrophysics, University of Delhi, Delhi 110007}

\date{Accepted XXX. Received YYY; in original form ZZZ}

\pubyear{2018}

\begin{document}

\date{Accepted ------------, Received ------------; in original form ------------}
\pagerange{\pageref{firstpage}--\pageref{lastpage}} \pubyear{}
\maketitle
\label{firstpage}

\begin{abstract}
We present dynamical status of the Galactic globular cluster NGC 6656 using spatial distribution of Blue Straggler Stars (BSSs). A combination of multi-wavelength high-resolution space and ground-based data are used to cover a large cluster region. We determine the centre of gravity ($C_{grav}$) and construct the projected density profile of the cluster using the probable cluster members selected from \textit{HST} and \textit{Gaia} DR2 proper motion data sets. The projected density profile in the investigated region is nicely reproduced by a single mass King model, with core ($r_{c}$) and tidal ($r_{t}$) radius as $75\arcsec.2$ $\pm$ $3\arcsec.1$ and $35\arcmin.6$ $\pm$ $1\arcmin.1$ respectively. In total, 90 BSSs are identified on the basis of proper motion data in the region of radius $625\arcsec$. An average mass of the BSSs is determined as $1.06$ $\pm$ $0.09$ $M_{\sun}$ and with an age range of 0.5 to 7 Gyrs. The BSS radial distribution shows a bimodal trend, with a peak in the centre, a minimum at $r \sim r_c$ and a rising tendency in the outer region. The BSS radial distribution shows a flat behaviour in the outermost region of the cluster. We also estimate $A^{+}_{rh}$ parameter as an alternative indicator of the dynamical status of the cluster and is found to be $0.038$ $\pm$ $0.016$. Based on the radial distribution and $A^{+}_{rh}$ parameter, we conclude that NGC 6656 is an intermediate dynamical age cluster.

\end{abstract}



\section{Introduction}
Globular clusters (GCs) are one of the oldest systems, where stellar interactions occur due to the high stellar density of the system. These stellar interactions lead to several dynamical processes, such as stellar collisions, core collapse, stellar mergers, two-body relaxation and mass segregation from equipartition of energy \citep{1997A&ARv...8....1M}. These dynamical processes result in several exotic populations, like low mass millisecond pulsars, cataclysmic variables, and blue straggler stars (BSSs) \citep{1995ARA&A..33..133B,2001ApJ...561..337F}. 
Among these exotic populations, BSSs are found in bulk and therefore play a crucial role in understanding the GC internal dynamics. 

The BSSs were discovered in 1953 by Sandage in the external regions of the GC M3. 
\cite{1953AJ.....58...61S} found that these stars appear as an extension of the main sequence and are hotter and luminous than the main sequence turnoff point. The actual nature of these peculiar objects is still not clear. Observational evidences \citep{1997ApJ...489L..59S} have shown that BSSs are more massive ($M$ $\sim$ 1.2 $M_{\sun}$) 
than average stars in the GCs ($M$ $\sim$ 0.3 $M_{\sun}$). Hence, they can be affected by dynamical friction, which segregates the BSSs towards the cluster centre \citep{2006ApJ...638..433F}. 

The construction of a complete sample of the BSSs has always been a challenging task. Because of stellar crowding in the central region and the dominant contribution of the luminous cool and bright giants (e.g, RGB, SGB), it is difficult to identify the BSSs in optical bands. However, in ultraviolet (UV) bands, RGB stars are among the faint ones, whereas BSSs are the brightest objects to be easily identified in this plane \citep{1991Natur.352..297P}. Therefore, with the help of UV observations taken with the \textit{Hubble Space Telescope} (\textit{HST}), it has become possible to study BSSs in the central regions of the dense clusters. 

The selection of BSSs is a difficult task on the basis of only photometric data. After the second \textit{Gaia} Data Release, \textit{Gaia} DR2, it has now become possible to separate the genuine BSSs from the field stars using proper motion information. This allows the study of BSS radial distribution from the very central out to large cluster region using \textit{HST} and wide-field imagers mounted on ground-based telescopes. Various studies have been done on the distribution of BSS in the literature. On the basis of the shape of observed BSS radial distribution, \citet[][]{2012Natur.492..393F} (hereafter F12) grouped GCs into three families;

In \textit{Family I}, the BSS radial distribution shows flat distribution, 
suggesting that dynamical friction has not yet segregated the BSSs located in the innermost cluster region. Various investigations for many GCs e.g., 
$\omega$ Centauri studied by \cite{2006ApJ...638..433F}, Palomar 14 by \cite{2011ApJ...737L...3B} and NGC 2419 by \cite{2008ApJ...681..311D} show flat BSS radial distribution and are classified as \textit{Family I} clusters. 

Several clusters show bimodality in their BSS radial distribution and are classified as \textit{Family II} clusters. The example of \textit{Family II} clusters are: M53 \citep{2008ApJ...679..712B}, NGC 6388 \citep{2008ApJ...677.1069D}, 47 Tuc \citep{2004ApJ...603..127F}, M5 \citep{2007ApJ...663..267L}, M55 \citep{2007ApJ...670.1065L}, NGC 6752 \citep{2004ApJ...617.1296S} and NGC 5824 \citep{2014ApJ...780...90S}. In these clusters bimodality is seen with a peak in the inner region, a clear minimum at ($r_{min}$), and an external rising trend. 
The $r_{min}$ tells the distance up to which the role of dynamical friction can be effectively seen. 

\textit{Family III} clusters show a monotonically decreasing radial distribution of BSS, with only a central peak followed by a rapid decline and no signs of an external upturn. Many GCs e.g., M80 $\&$ M30 studied by F12, M79 by \cite{2007ApJ...663.1040L} and M75 by \citet[][]{2012ApJ...748...91C} have shown unimodal behavior in BSS radial distribution. In these clusters, the dynamical friction has already moved the most distant BSS toward the cluster centre. 

As mentioned in F12, these different families correspond to the different dynamical age of the clusters. The Family I, II and III are termed as dynamically young, dynamically intermediate and the dynamically old clusters respectively. 

To measure the segregation level of BSSs, \cite{2016ApJ...833..252A} proposed a new parameter ($A^{+}$), which is defined as the area between the cumulative radial distribution curves of BSSs ($\phi_{BSS}$(x)) and the reference population ($\phi_{REF}$(x)):

\begin{equation}
 A^{+}(x) = \int^{x}_{x_{min}} \phi_{BSS}(x') - \phi_{REF}(x') dx'
\end{equation}

where x (= log($r/r_{h}$)) is the logarithmic distance from the cluster centre in units of the half-mass radius $r_{h}$ of the cluster. For meaningful cluster to cluster comparison, \citet[][]{2016ApJ...833L..29L} (hereafter L16) defined the measure of this parameter (as $A^{+}_{rh}$) within one half-mass radius ($r_{h}$). They found a tight correlation between $r_{min}$ and $A^{+}_{rh}$ for a sample of 25 Galactic GCs. The value of $A^{+}_{rh}$ is, therefore, an alternative indicator to understand the dynamical status of the cluster.

In this paper, we present the dynamical status of the Galactic GC NGC 6656 (M22) using radial distribution of BSS population. NGC 6656 ($\alpha_{J2000}$ = $18^{h}$ $36^{m}$ $23^{s}.94$, 
$\delta_{J2000}$ = $-23\degr$ $54\arcmin$ $17\arcsec$.1 , \citet[][]{2010arXiv1012.3224H}\footnote{\url{http://physwww.physics.mcmaster.ca/~harris/mwgc.dat}} (hereafter HA10) is one of the nearest 
globular cluster at a distance of 3.2 kpc from the Sun. The core radius of NGC 6656 is $\sim$ 1.3 pc and this is a low density cluster ($\textless$ $10^4$ $M_{\sun}$ $pc^{-3}$). \cite{2016ApJ...827...12B} studied the BSS kinematic profile of 19 GCs including NGC 6656 using \textit{HST} proper-motion catalogue  and they identified 34 BSSs in the central region of NGC 6656. They calculated the average mass of BSSs as $1.49^{+0.47}_{-0.28}$ $M_{\sun}$. Present analysis of the BSS is performed in larger 
region (625$\arcsec$ in radius) of the cluster. In addition to this BSSs are selected from the \textit{HST} proper motion catalogue in the central 
region and \textit{Gaia} DR2 catalogue in the outer region of the cluster. 

\section{DATA SETS AND ANALYSIS } 
\subsection{The data sets}

In order to select the BSSs in the entire sample, we need high resolution (HR) data set in the central region and wide field (WF) data set in the outer region of the cluster. Therefore, we used \textit{HST} data in the centre and ground-based CCD data
taken from 2.2-m ESO/MPI telescope in the outer region of the cluster. 

For the HR data set, we used the astro-photometric catalogue\footnote{\url{http://www.astro.uda.cl/public_release/globularclusters41.html}} provided by \citet{2017AJ....153...19S}. The data set used in this catalogue are taken from the \textit{Hubble Space Telescope UV Legacy Survey of Galactic Globular Clusters} (GO-13297) and from two other projects GO-12311 and GO-12605 \citep{2015AJ....149...91P}. The photometric data in this catalogue are in \textit{F275W}, \textit{F336W} and \textit{F438W} filters which are taken from Wide Field Camera 3 (WFC3) through Ultraviolet-Visible (UVIS) channel. The WFC3/UVIS camera consists two chips, each of $4096\times2051$ pixels with a pixel scale of $0\arcsec.0395$ $pixel^{-1}$ provides a field of view (FOV) of $\sim$ $162\arcsec\times162\arcsec$.
This catalogue also provides photometric data in \textit{F606W} and \textit{F814W} filters taken from ACS/WFC (GO-10775, PI-Sarajedini). The ACS/WFC camera provides a total FOV of $\sim$ $202\arcsec\times202\arcsec$ with a pixel scale of $0\arcsec.05$ $pixel^{-1}$. The respective circular FOV is shown in the upper right panel of Fig. 1. In addition to the photometric data, the catalogue also contains relative proper motion information of stars which is calculated by using the ACS and WFC3 UV positions as the first and second epoch, respectively.

For WF sample, we used archival images\footnote{\url{http://archive.eso.org/eso/eso_archive_main.html}} taken with the Wide-Field Imager (WFI) mounted at 2.2-m ESO/MPI telescope (La Silla, Chile). A set of wide-field \textit{B}, \textit{V}, \textit{I}, and \textit{U} images are used to sample the BSSs over larger FOV. The WFI data set in \textit{B}, \textit{V}, \textit{I} were observed on 1999 May 15, and \textit{U} data set was observed on 2012 April 20. The aim of selecting \textit{U} filter was to fill the gaps between the CCD chips. The WFI consists of eight CCD chips placed together (each composed of 2048 $\times$ 4096, with a pixel scale $\sim 0\arcsec.238$ $pixel^{-1}$), with a total FOV of $\sim$ $34\arcmin$ $\times$ $33\arcmin$ as shown in Fig. 1.

\begin{figure}
	\includegraphics[width=\columnwidth]{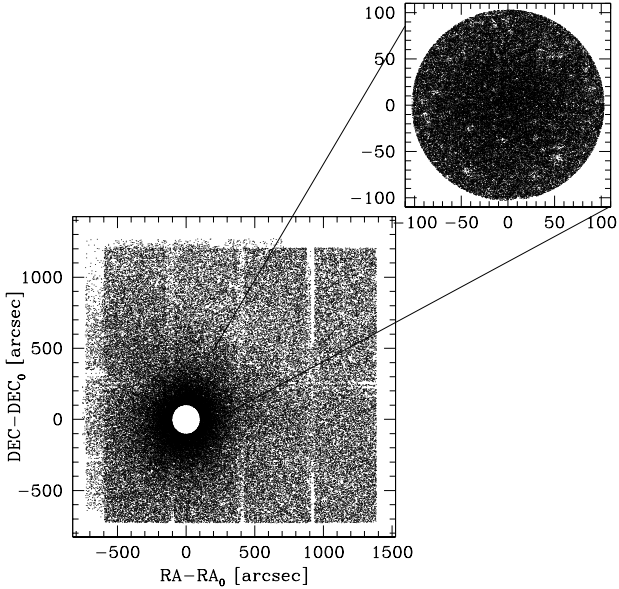}
    \caption{The wide-field data set covering the large cluster region. The map of all the stars found in the WF images are plotted with respect to $C_{grav}$ and the open region located towards the cluster centre is taken from the HR sample. The map of high-resolution data set is shown on the top right of the figure.}		\label{fig:general}
\end{figure}

\subsection{The data reduction and calibration of 2.2-m ESO data set} 
The WF images are reduced with the procedures explained in \citet[][]{2006A&A...454.1029A} (hereafter A06). 
The pre-processing are done using \textit{MSCRED} package under IRAF. As mentioned in A06, the shape of the PSF in the WF images varies significantly with the position. Therefore, an array of 9 PSFs per CCD chip ($3\times3$) is constructed to minimize the spatial variability down to 1 per cent. The PSFs are constructed from an empirical grid of the size of a quarter pixel. Each PSF is represented by a 201 $\times$ 201 grid, centred at (101,101) and extend out to a radius of 25 pixels. 

In order to find the instrumental magnitudes and positions from the array of PSFs, an automated code described in A06 is used. The code starts by finding the brightest stars and goes deeper into the fainter stars. Also, to minimize the effect of geometric distortion, the distortion solution derived in A06 is used. However, the geometric distortion solution, particularly for $U$ is not appropriate for astrometry (A06) and the astrometric solutions can be affected due to systematic errors.

To transform the averaged \textit{X}, \textit{Y} coordinates to the corresponding right ascension ($\alpha_{J2000}$) and declination ($\delta_{J2000}$) we used \textit{CCMAP} and \textit{CCTRAN} tasks available in IRAF. An accuracy of $\sim0\arcsec.1$ is obtained in RA and DEC transformation. The $U$, $B$, $V$ and $I$ instrumental magnitudes (exposure time and airmass corrected) of WF images are converted into the \textit{F336W}, \textit{F438W}, \textit{F606W}, and \textit{F814W} VEGAMAG photometric system by using the transformations derived on the basis of about thousand common stars between HR and WF data sets. Only stars brighter than 20 mag in the $F606W$ filter are used for calibration purposes.

In order to derive the photometric zero-points and colour terms, we used the following transformation equations;\\

$F336W_{std}$ = $U_{ins}$ + $C_{u}$ $\ast$ ($U_{ins} - B_{ins}$) + $Z_{u}$

$F438W_{std}$ = $B_{ins}$ + $C_{b}$ $\ast$ ($B_{ins} - V_{ins}$) + $Z_{b}$

$F606W_{std}$ = $V_{ins}$ + $C_{v}$ $\ast$ ($V_{ins} - I_{ins}$) + $Z_{v}$

$F814W_{std}$ = $I_{ins}$ + $C_{i}$ $\ast$ ($V_{ins} - I_{ins}$) + $Z_{i}$\\

where instrumental magnitudes and standard magnitudes are represented by the subscript ``ins'' and ``std'', respectively. In the above equations, $C_{u}$, $C_{b}$, $C_{v}$, and $C_{i}$ denote the colour coefficients, while $Z_{u}$, $Z_{b}$, $Z_{v}$, and $Z_{i}$ are the zero-points. The corresponding values of the colour coefficients are 0.37, 0.47, -0.30, and 0.028 and zero-points are 21.09, 24.77, 24.06, and 23.44 mag respectively. 
 
\section{Field star decontamination} \label{sec:conta}

\begin{figure}
	\includegraphics[width=\columnwidth]{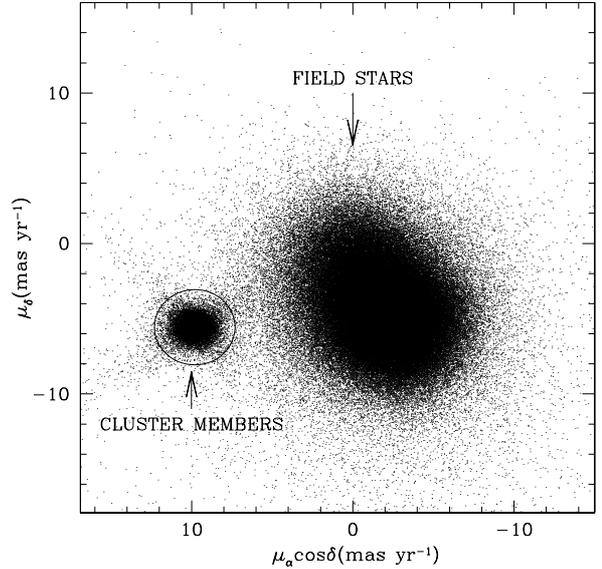}
	\caption{The Vector-point diagram (VPD) for the stars having the proper motion information in the \textit{Gaia} DR2 catalogue. The criteria for selection of cluster members is indicated by a circle of radius 3.0 mas $yr^{-1}$ around the cluster centre (9.81, -5.57) mas $yr^{-1}$.}
	\label{fig:general}
\end{figure}

The photometric data sets are contaminated by the presence of field stars. The field star contamination can be reduced by using proper motion information of the stars. \cite{2017AJ....153...19S} contains relative proper motions information for the stars common in ACS and WFC3 FOV. An analysis of the VPD shown in 
\cite{2017AJ....153...19S} exhibits that the stars lying outside the radius 0.35 pixel can be considered as field stars. We have adopted the same criteria as discussed in \cite{2017AJ....153...19S} for selection of members for the HR sample.

The \textit{Gaia} DR2 catalogue contains the astrometric and photometric information for all the stars down to $G$ $\sim$ 21 mag ~\citep[][]{2016A&A...595A...1G,2018A&A...616A...1G}. The \textit{Gaia} DR2 proper motion data are used to select the members outside the HR sample. The VPD for all the stars except the HR sample is shown in Fig. 2. The stars plotted in the VPD are having an accuracy of $\leq$ 1 mas $yr^{-1}$ in both the 
proper motion directions. From the VPD we can see that cluster stars are clearly separated from field stars. The centres of the cluster and field star distributions are (9.81,-5.57) and (-1.72, -4.35) mas $yr^{-1}$ respectively. We define the selection criteria for cluster members located within the circle of radius 3.0 mas $yr^{-1}$ around the centre of cluster distribution as shown in Fig. 2. For further analysis, we used only those stars which are lying within the radius of 3.0 mas $yr^{-1}$.

\section{Results and Discussions} \label{sec:floats}
\subsection{Centre of Gravity} \label{subsec:grav}
Stellar evolutionary theories say that stellar luminosities are not directly proportional to stellar masses in a cluster \citep{1995MNRAS.276..739M}. The estimation of 
geometrical centre ($C_{grav}$) for the cluster may differ significantly from previously derived centre of luminosity ($C_{lum}$) values using surface brightness distribution. By considering the centre reported in HA10 as initial guess, the centre of gravity $C_{grav}$ is estimated by taking the average of $\alpha$ and $\delta$ of stars for HR data set. Using the iterative method 
described in \citet[][]{2012ApJ...748...91C}, we constructed four circular areas with different radii ($10\arcsec$, $11\arcsec$, $12\arcsec$ and $13\arcsec$) and for each of these radii, three different magnitude limits 
($m_{F606W}$= 20.5, 21 and 21.5 mag) are consider to reduce the effect of incompleteness. We estimated $C_{grav}$ as $\alpha_{J2000}$ = $18^{h}$ $36^{m}$ $24^{s}.0$ and 
$\delta_{J2000}$ = $-23\degr$ $54\arcmin$ $16.\arcsec4$ by averaging the values of $C_{grav}$ computed over all the 12 different combinations. Our calculated $C_{grav}$ value differs with $0\arcsec.5$ in $\alpha$ and $\delta$ from the value estimated by \cite{2010AJ....140.1830G}, using density contour method.

\subsection{Projected density profile} \label{subsec:dens}
To construct the Projected density profile (PDP) of the cluster, we made a catalogue of stars with \textit{HST} in the centre and \textit{Gaia} DR2 in the outer region. We considered \textit{Gaia} DR2 stars because the catalogue is complete up to \textit{G} = 18 mag ~\citep[][]{2018A&A...616A..17A}. We also transformed the \textit{HST} \textit{F606W} to \textit{Gaia} \textit{G}-band magnitude using the relation derived in ~\citet[][]{2010A&A...523A..48J}.

In order to obtain the structural parameters of the cluster, we constructed the PDP of NGC 6656, starting from $C_{grav}$ out to $\sim$ $36\arcmin$ (tidal radius). We divided the area into 19 concentric annuli with varying radii centred on $C_{grav}$, and each annulus is divided into four equivalent sub-sectors. The density estimation for each sub-sector is done by counting the stars present in the region divided by the area of sub-sector. The resulting density for each annulus is estimated by averaging the densities of the sub-sectors and standard deviation in the average is considered as the uncertainty in the annulus densities. The radius of each annulus is considered as the midpoint value of the corresponding radial bin.

\begin{figure}
	\includegraphics[width=\columnwidth]{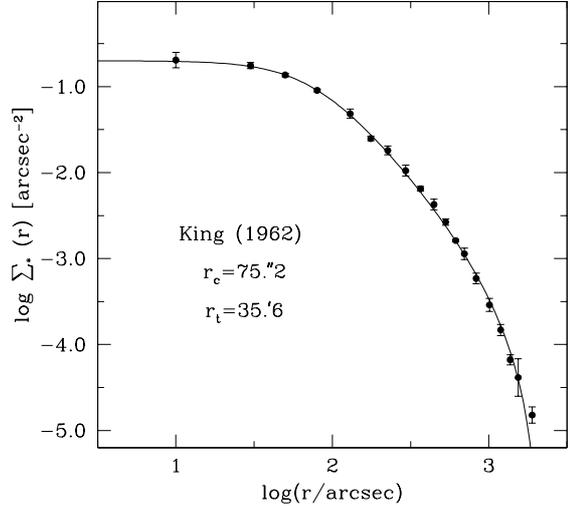}
	\caption{Observed projected density profile, plotted over $\sim$ $36\arcmin$ for the cluster NGC 6656. The density profile is nicely reproduced by an isotropic single-mass King model with the parameters shown in the box. Continuous line shows the King (1962) profile.}
	\label{fig:general}
\end{figure}

The resultant PDP of NGC 6656 is shown in Fig. 3. The PDP is nicely fitted with an isotropic single-mass King model given by ~\citet[][]{1962AJ.....67..471K} . The fitting provides the core ($r_{c}$) and tidal ($r_{t}$) radii as $75\arcsec.2$ $\pm$ $3\arcsec.1$ and $35\arcmin.6$ $\pm$ $1\arcmin.1$ respectively. In HA10 and ~\citet[][]{2014A&A...572A..30K}, the ($r_{c}$) and 
($r_{t}$) are listed as $r_{c}$ $\simeq$ $79\arcsec.8$ and $r_{t}$ $\simeq$ $31\arcmin.9$. \cite{1995AJ....109..218T} have also estimated $r_{c}$ and $r_{t}$ as $85\arcsec.1$ and $28\arcmin.9$ respectively, using surface brightness profile of the cluster. The present estimate of $r_{c}$ and $r_{t}$ are quite different with the previous studies. However, the current estimate is based on proper motion data and considered to be more reliable than the earlier investigations.

\subsection{Selection of BSSs and reference populations} \label{subsec:def}
To study the BSS radial distribution, the first step is to select genuine BSSs and reference populations (HB or GB (RGB+SGB)) carefully. For this purpose, we used the selected cluster members as described in Sec. 3.

\subsubsection{The BSS population}
We adopted UV-CMD ($m_{F275W}$, $m_{F275W}-m_{F336W}$) as our primary selection criteria for selecting the BSSs. In this plane, BSSs are easily distinguishable from cooler giants (RGB or AGB). 
As shown in Fig. 4, the BSS populations follow a vertical sequence in the UV-CMD. The contamination from the sub-giant branch and MS-TO is minimized by adopting $m_{F275W}$=18.65 mag as the limiting magnitude of the BSS selection box. A total of 26 BSSs are identified in the inner sample ($r$ $\leq$ $104\arcsec$), using UV-CMD. These BSSs are also checked for membership as described in Sec. 3 and found to be genuine members of the cluster.

The BSSs once selected from the UV-CMD are replotted in the Optical-CMD ($m_{F606W}$, $m_{F606W}-m_{F814W}$) to define the size of the selection box in optical band. Based on the position of 26 BSSs in Optical-CMD, we defined the selection box criteria (15.65 $\textless$ $m_{F606W}$ $\textless$ 17.12 and 0.36 $\textless$ $m_{F606W}-m_{F814W}$ $\textless$ 0.65) as shown in left panel of Fig. 5. Six more BSSs in the ACS FOV, not covered by the WFC3 FOV are identified in the inner sample using Optical-CMD. Based on the member selection norms, they are bonafide members of the cluster. Finally, a total of 32 BSSs are identified in the inner sample. \cite{2016ApJ...827...12B} found 34 BSSs in the region of $202\times$202 $arcsec^{2}$ using ($m_{F814W}$, $m_{F606W}-m_{F814W}$) CMD. The investigated area by \cite{2016ApJ...827...12B} is more than the present analysis.

\begin{figure}
	\includegraphics[width=\columnwidth]{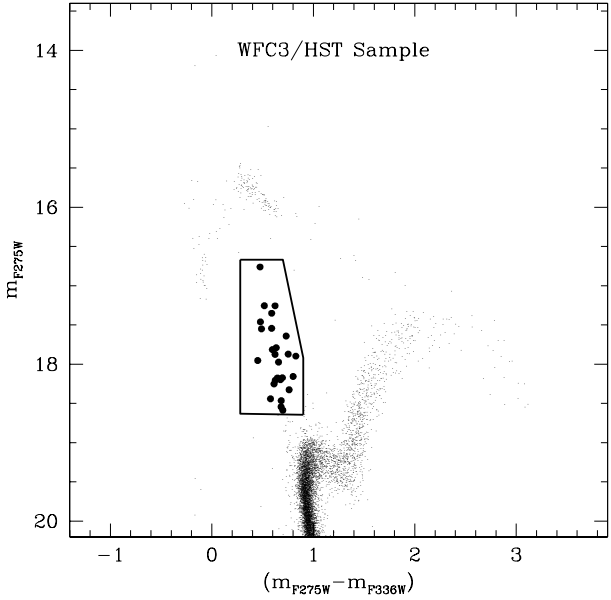}
	\caption{The selection criteria for the BSSs in UV-CMD ($m_{F275W}$, $m_{F275W}-m_{F336W}$) for the HR sample is shown with box. The selected BSSs are shown with filled dots.}
	\label{fig:general} 
\end{figure}

\begin{figure}
	\includegraphics[width=\columnwidth]{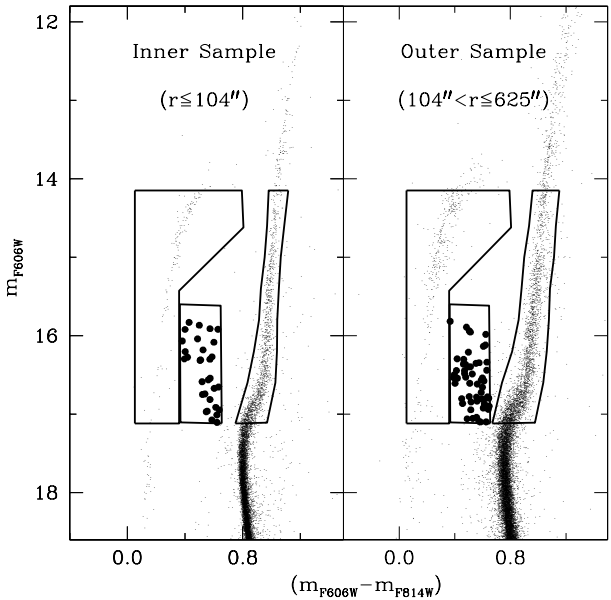}
	\caption{$left$ $panel$: the Optical-CMD ($m_{F606W}$, $m_{F606W}-m_{F814W}$) of the inner sample ($r$ $\leq$ $104\arcsec$). $right$ $panel$: the Optical-CMD of the outer sample ($104\arcsec$ $\textless$ $r$ $\leq$ $625\arcsec$). The selection boxes for selecting the BSSs, HB, and GB are also shown. The BSSs are represented by filled dots.}  
	\label{fig:general}
\end{figure}

For the outer sample ($104\arcsec$ $\textless$ $r$ $\leq$ $625\arcsec$), we adopted the same selection box as used for the inner sample and shown in the right panel of Fig. 5. In this way, we found 58 BSSs in the outer sample. Therefore, a total of 90 BSSs are identified from the entire (inner+outer) sample on the basis of photometric and proper motion data.
 
\subsubsection{The selection of reference populations}
In order to carry out a qualitative study of BSS in terms of radial distribution and specific frequency, the definition of the reference population is very necessary. The reference population should show a non-peculiar radial trend within the cluster. According to \citet{2007ApJ...670.1065L} and \citet{1988ARA&A..26..199R}, the number of stars in any post-MS stage is proportional to the duration of the evolutionary phase itself. The specific frequencies ($N_{HB}/N_{GB}$) are expected to be constant and equal to the ratio between the evolutionary timescales of the HB and GB phase. Therefore, we considered these branches as reference populations. We set the same fainter magnitude limit for BSSs and reference populations so that our selection is not affected by the completeness of the sample. For the selection of the reference population, we used Optical-CMD as shown in Fig. 5. We made a box around the HB population in such a way that it should contain most of the HB stars. In this way a total of 428 (119 in the inner sample and 309 in the outer sample) HB reference stars are identified. 

For GB population selection, we defined a mean ridge-line by taking the average in colour ($m_{F606W}-m_{F814W}$) in the interval of 0.4 mag in $m_{F606W}$ magnitude. For each sample (inner and outer), we used a $3\sigma$ selection box criteria defined from the mean ridge line. Here, $\sigma$ is the standard deviation in the mean colour. This is done to reduce the effects of differential reddening and error in photometric calibration. In this way, we found 1233 and 2779 GB reference stars in the inner and outer sample respectively.

Finally, we found 428 HB stars and 4012 GB stars as reference population in the entire sample.

\begin{table}
	\centering
	\caption{The frequency distribution of BSSs in different mass intervals}
	\label{tab:example_table}
	\begin{tabular}{lc} 
		\hline
		\hline
		Mass range & Frequency\\
		\hline
		0.90-0.95 & 3\\
		0.95-1.00 & 24\\
		1.00-1.05 & 18\\
		1.05-1.10 & 14\\
		1.10-1.15 & 12\\
		1.15-1.20 & 5\\
		1.20-1.25 & 9\\
		1.25-1.30 & 4\\
		1.30-1.35 & 1\\
		\hline
	\end{tabular}
\end{table}

\subsection{The BSS Mass Distribution}
The location of BSSs in Optical-CMD suggest that they are massive in comparison to the normal stars in the GCs. Their masses are estimated by comparing their location in the CMD with theoretical isochrones taken from \cite{2002A&A...391..195G}. Fig. 6 shows the set of theoretical isochrones fitted in the population of 90 BSSs with an age range of 0.5 to 7 Gyrs, in a step of 0.5 Gyr. The metallicity ($[Fe/H]$ = $-$1.72) and the distance modulus ($(m-M)_{V}$ = 13.60) values of the cluster are adopted from HA10 catalogue. The projection of the BSSs on the nearest isochrone is used to derive their masses. In this way, BSS masses are estimated in a range of 0.90$-$1.35 $M_{\sun}$. The frequency distribution of BSSs in the different mass interval is listed in Table 1. It is clear from the table that the maximum number of BSS stars are in the range of 0.95-1.00 $M_{\sun}$. The average mass of BSS is found to be $1.06$ $\pm$ $0.09$ $M_{\sun}$.

\begin{figure}
	\includegraphics[width=\columnwidth]{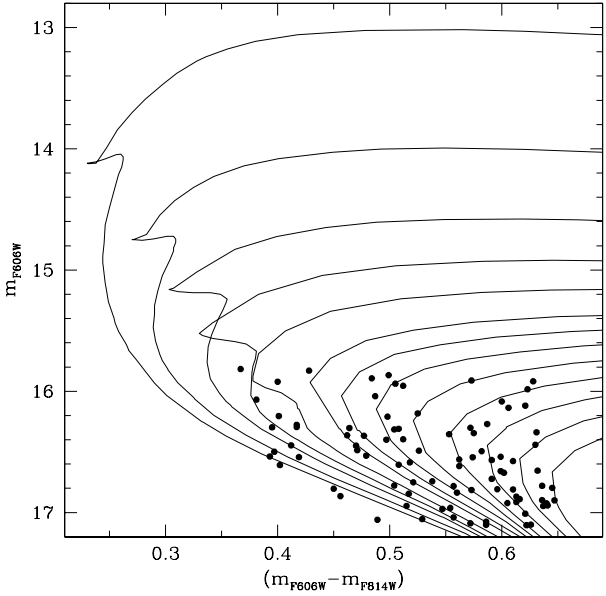}
	\caption{The Optical-CMD for 90 BSS identified in the present analysis. The continuous lines are 
the theoretical isochrones taken from Girardi et al. (2002) and fitted to the BSS.}
	\label{fig:general}
\end{figure}

\cite{2007ApJ...663.1040L} has obtained the average mass of the BSS as 1.2 $M_{\sun}$ in the cluster NGC 1904. \cite{2016ApJ...827...12B} has provided the BSS kinematic profile for 19 GCs using ACS/\textit{HST} sample and estimated the average mass as $1.49^{+0.47}_{-0.28}$ $M_{\sun}$ for the BSSs lying near the centre of the cluster NGC 6656. The present estimate of the average mass of BSSs in NGC 6656 is similar to the value derived by previous authors. 

\subsection{The radial distribution of BSS, GB, and HB} \label{subsec:bss}

In this section, we present the radial distribution of BSS, GB, and HB populations. The cumulative radial distribution of BSS, GB, and HB are shown with continuous, dashed and dotted lines respectively in Fig. 7. A comparison of the BSS cumulative radial distribution with GB and HB distribution indicates that BSSs are more centrally concentrated than reference populations. To get the more clear picture about the distributions we performed the Kolmogorov-Smirnov test\footnote{\url{http://www.physics.csbsju.edu/stats/KS-test.html}}. The probabilities with which the BSSs have a different radial distribution than the HB and GB populations are 98.3 and 99.9 per cent respectively. This shows that BSS population is extracted from different parent population than the reference populations.  
\begin{figure}
	\includegraphics[width=\columnwidth]{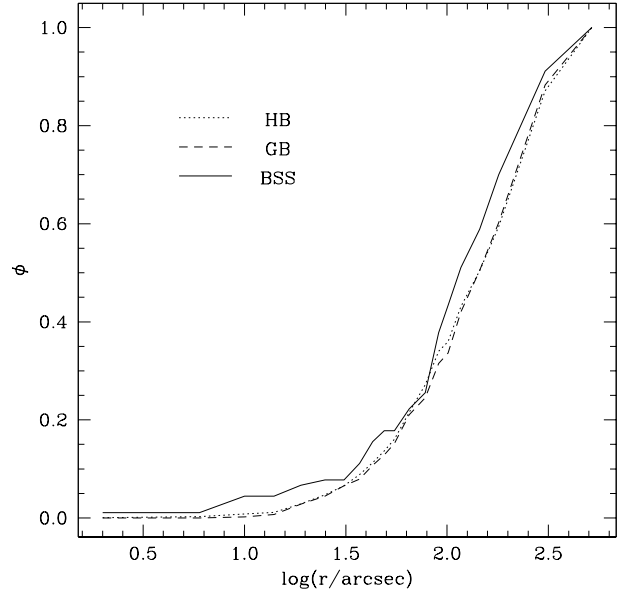}
	\caption{Cumulative radial distribution plot of the BSSs (continuous line), HB stars (dotted line) and GB stars (dashed line) with respect to $C_{grav}$ of the cluster.}								\label{fig:general}
\end{figure} 

In order to investigate the radial distribution behavior of BSSs, GB, and HB further, the cluster region is divided into seven concentric annuli centred on $C_{grav}$. We counted the number of BSSs, GB and HB populations in each bin and listed in Table 2 as $N_{BSS}$, $N_{GB}$ and $N_{HB}$. We then computed the specific frequencies $F^{BSS}_{GB}$ = $N_{BSS}$/$N_{GB}$,  
$F^{BSS}_{HB}$ = $N_{BSS}$/$N_{HB}$ and $F^{HB}_{GB}$ = $N_{HB}$/$N_{GB}$. These Specific frequencies are plotted with respect to the radius in Fig. 8. The specific frequency $F^{HB}_{GB}$ shows a flat behavior across the entire field of investigation of the cluster. On the other hand, the specific frequencies of BSS show a bimodal trend with both the reference 
populations. A peak in the centre, a minimum at $r \sim r_{c}$ and a rising trend in the outer region of the cluster are present. A flattening is also seen in the outermost region of the cluster. 
 
\begin{figure}
	\includegraphics[width=\columnwidth]{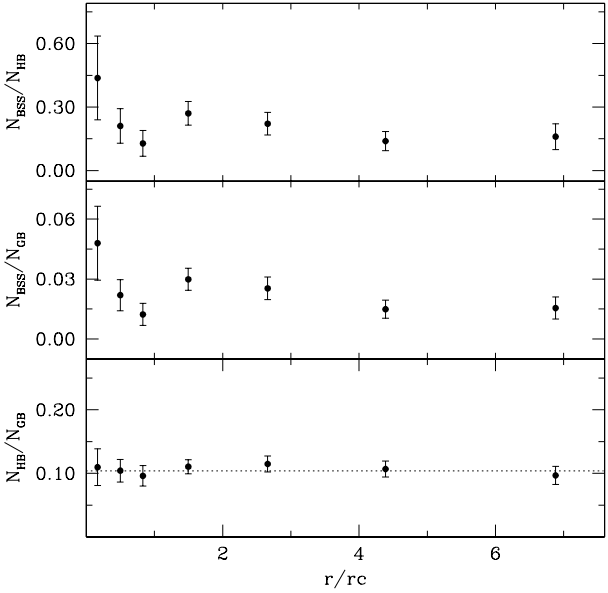}
	\caption{The radial distribution of the specific frequencies $F^{BSS}_{HB}$ (top), $F^{BSS}_{GB}$ (middle), and $F^{HB}_{GB}$ (bottom) plotted with respect to distance from the cluster centre normalized by the core radius. The dotted line shows the mean of all the points plotted in the bottom panel.}
	\label{fig:general}
\end{figure}

\begin{table}
	\centering
	\caption{The log of the Number counts for BSSs and reference populations} 
	\label{tab:decimal}
	\begin{tabular}{ccccc}
		\hline
		\hline
		Radial bin (in arcsec) & $N_{BSS}$ & $N_{HB}$ & $N_{GB}$ & $L_{samp}/L^{tot}_{samp}$ \\
		\hline
		0 - 25 & 7 & 16 & 146 & 0.03 \\
		25 - 50 & 8 & 38 & 365 & 0.07 \\
		50 - 75 & 5 & 39 & 406 & 0.09 \\
		75 - 150 & 30 & 111 & 1005 & 0.24 \\
		150 - 250 & 21 & 95 & 828 & 0.22 \\
		250 - 410 & 11 & 79 & 740 & 0.20 \\
		410 - 625 & 8 & 50 & 517 & 0.14 \\
		\hline
	\end{tabular}
\end{table}

To recheck the bimodality in BSS radial distribution, we estimated the doubly normalized ratio for BSS. It is defined as the number of BSSs observed in a region to the total number of BSSs divided by the fraction of light sampled in the same region with respect to the total measured luminosity (in $L_{\sun}$) \citep{1993AJ....106.2324F}. It is written as,

\begin{equation}
R_{BSS} = \frac{N_{BSS}/N^{tot}_{BSS}}{L_{samp}/L^{tot}_{samp}}
\end{equation}

We also computed this ratio for reference stars. By adopting the parameters obtained in Sec. 4.2, we calculated the sampled luminosity for each annulus by integrating the isotropic single-mass 
King profile and scaled it to the area covered between the annular regions. We considered Poisson error for the different populations and luminosities. The error in double normalized ratios is considered as the propagation of errors \citep{2004ApJ...617.1296S}. The computed luminosity ratios in each annulus are listed in Table 2.

Fig. 9 shows the variation of double normalized ratios of BSS ($R_{BSS}$), HB ($R_{HB}$) and GB ($R_{GB}$) populations with the radial distance normalized to $r_{c}$. The upper panel represents $R_{BSS}$ and $R_{HB}$, while lower panel represents $R_{BSS}$ and $R_{GB}$ with respect to $r/r_{c}$. The value of $R_{BSS}$ 
show a peak in the centre, a dip at $r$ $\sim$ $r_c$ and an external rising trend followed by flattening in the outskirt of the cluster. The nature of this plot is very similar to Fig. 8. The resulting radial distributions of $R_{GB}$ and $R_{HB}$ follow the cluster luminosity and is almost a constant value ($\sim$ 1) as expected from the theory of stellar evolution \citep{1988ARA&A..26..199R}. 

\subsubsection{Discussion on the BSS radial distribution}

An analysis of Fig. 8 and 9 shows that BSS radial distribution in NGC 6656 is bimodal. Dynamical friction plays a very crucial role in shaping the BSS radial distribution. It segregates the massive objects like BSS that are rotating close to the centre of the cluster. As a result, a peak in the centre and a dip in smaller radii is visible in the BSS radial distribution. The BSSs located in the outskirts of the cluster have not yet been influenced by the action of dynamical friction and hence show a flat distribution. The BSS radial distribution found for NGC 6656 is similar to the previously studied GC M5 \citep{2007ApJ...663..267L}. The features in the BSS radial distribution can be used as a measure of dynamical age. Based on the shape of BSS radial distribution, this cluster seems to be dynamically intermediate age cluster. 

\begin{figure}
	\includegraphics[width=\columnwidth]{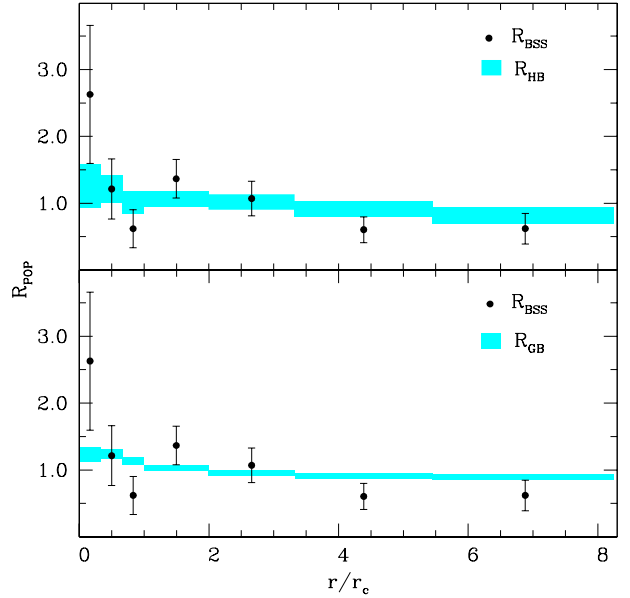}
	\caption{The BSSs radial distribution (filled circles) and the reference populations with double normalized ratios are plotted with respect to the distance from the cluster centre, normalized to $r_{c}$. The shaded portion in the upper and lower panel shows the distribution of HB and GB population respectively, which is almost constant ($\sim$ 1). The width of the shaded regions represent the error bars.}
	\label{fig:general}
\end{figure}

\subsection{$A^{+}_{rh}$ determination and dynamical status of the cluster} \label{subsec:dyn}

In the previous section, we discussed the dynamical state of the cluster based on BSS radial distribution in the region of the cluster. Despite several observational confirmations, it has not been possible to reproduce the BSS radial distribution profile from Monte Carlo and $N$-body simulations of GCs. Also, the bimodality seems to be unstable and temporary feature \citep[see][]{2015ApJ...799...44M,2017MNRAS.471.2537H}. Therefore, to reconfirm the dynamical age, we estimated $A^{+}_{rh}$ parameter proposed by \cite{2016ApJ...833..252A}, as a measure of the dynamical status of the cluster. $A^{+}_{rh}$ parameter is defined as the area between the curves of cumulative radial distribution of reference and BSSs populations over the half-mass radius ($r_{h}$). The value $r_{h}$ $=$ $201\arcsec.6$ is taken from HA10.  

\begin{figure}
	\includegraphics[width=\columnwidth]{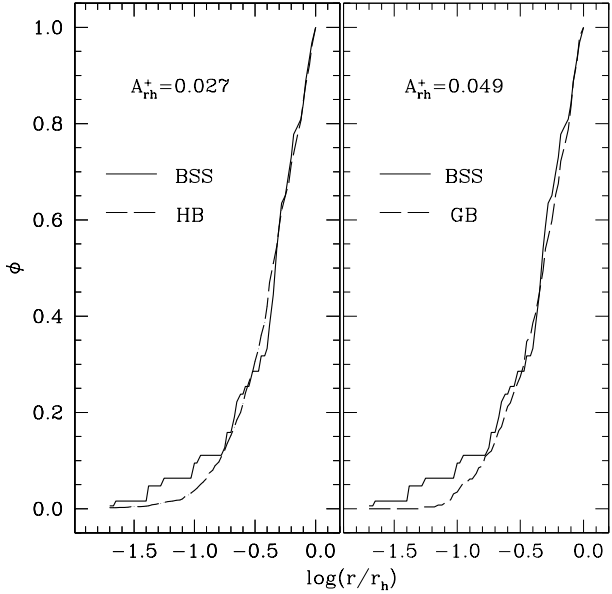}
	\caption{\textit{Left panel}: cumulative radial distribution of BSSs and reference population (HB) plotted over one half-mass radius ($r_{h}$). \textit{Right panel}: cumulative radial distributions of BSSs and GB as reference population over one $r_{h}$. The curve of reference populations are shown with a dashed line and the BSSs is shown with a continuous line.}
	\label{fig:general}
\end{figure}

In Fig. 10, we plot the cumulative radial distributions of BSSs and HB in the left panel and BSSs and GB in the right panel. Using these distributions the values of $A^{+}_{rh}$ are estimated as 0.027 and 0.049 for HB and GB sample respectively. The average value of $A^{+}_{rh}$= $0.038$ $\pm$ $0.016$. As discussed in L16, $A^{+}_{rh}$ can be adopted as an alternative indicator for measuring the level of dynamical evolution experienced by the cluster from the beginning. They have shown that $A^{+}_{rh}$ can be related to core relaxation time ($t_{rc}$/$t_{H}$). Here, $t_{H}$ = 13.7 Gyr is the age of the Universe and $t_{rc}$ = 0.34 Gyr is taken from HA10. We plotted $A^{+}_{rh}$ versus log($t_{rc}$/$t_{H}$) in lower panel of Fig. 11. The values of $A^{+}_{rh}$ and $t_{rc}$ for cluster NGC 6656 is shown with the filled circle and these values for other clusters are adopted from L16 and shown with empty circles. As discussed in L16, a decreasing trend of $A^{+}_{rh}$ with relaxation time is seen. It is clear from the plot that the cluster NGC 6656 follows the trend.  

In the framework of the \textit{empirical dynamical clock} relation defined in F12, the position of $r_{min}/r_{c}$ can be used as an indicator to measure the extent of dynamical evolution of the cluster. Therefore, we plot $r_{min}$/$r_{c}$ against $t_{rc}$/$t_{H}$ in the upper panel of Fig. 11. Filled circle represents the cluster NGC 6656. We adopted the value of $t_{rc}$, $r_{min}$ and $r_{c}$ for other clusters from F12. The filled triangles are dynamically old clusters while open circles are the dynamically intermediate age clusters. The dynamically
young clusters are plotted as lower-limit arrows at $r_{min}$ $\sim$ 0.1. An inspection of this figure exhibits that the cluster NGC 6656 lies in the region of dynamically intermediate age clusters. Hence, we suggest that this is an intermediate dynamical age cluster of \textit{Family II} classification.

Therefore, BSS radial distribution and $A^{+}_{rh}$ parameter indicates that NGC 6656 is a dynamically intermediate age cluster.

\begin{figure}
	\includegraphics[width=\columnwidth]{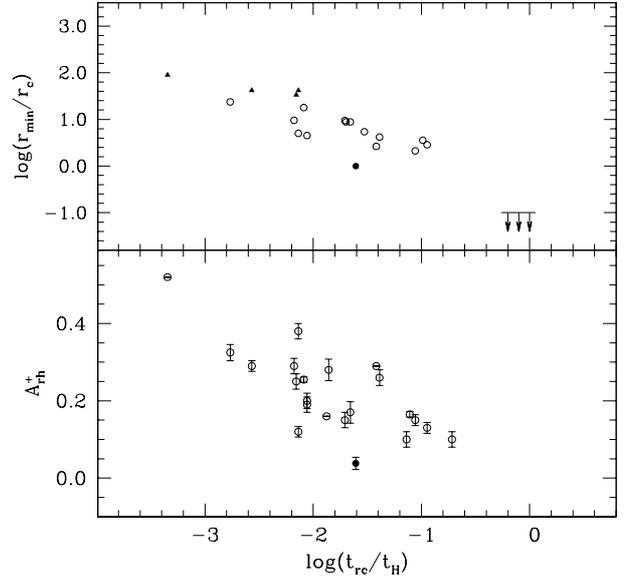}
	\caption{$Upper$ $panel:$ The plot shows the \textit{empirical dynamical clock} relation defined in F12, as a function of $r_{min}$/$r_{c}$ and $t_{rc}$/$t_{H}$. The open circles show the position of dynamically intermediate clusters and filled triangles for the dynamically old clusters. The dynamically young systems with the lower-limit arrows at $r_{min}$ = 0.1. The position of NGC 6656 is marked as the filled circle. $Lower$ $panel:$ show the position of NGC 6656 as the filled circle and other clusters are marked as open circles.}
	\label{fig:general}
\end{figure}

\section{SUMMARY AND CONCLUSIONS} \label{sec:summary}       
In this paper, a combination of multi-wavelength high-resolution space and ground-based data are used to probe the dynamical status of NGC 6656 based on BSS radial distribution. The BSS and reference populations are selected from both the photometric and kinematic data. 
The important findings of the present analysis are the following:
 
\begin{enumerate}

\item The centre of gravity, $C_{grav}$ of the cluster is determined as
$\alpha_{J2000}$ = $18^{h}$ $36^{m}$ $24^{s}.0$ and $\delta_{J2000}$ = $-23\degr$ $54\arcmin$ $16\arcsec.4$, with an accuracy of 
$\sim$ $0.05\arcsec$ in both $\alpha$ and $\delta$ using HR data set. The derived PDP can be nicely reproduced by an isotropic single-mass King model and provides $r_{c}$ = $75\arcsec.2$ $\pm$ $3\arcsec.1$ and $r_{t}$ = $35\arcmin.6$ $\pm$ $1\arcmin.1$. The stars used for 
the determination of $C_{grav}$ and PDP are selected using proper motion data.     

\item We identified a total of 90 BSSs in the entire sample, with 32 BSSs lying in the inner sample and 58 BSSs in the outer sample. These BSSs are 
selected using proper motion data taken from \textit{HST} and \textit{Gaia} DR2 catalogues. We estimated an average BSSs mass as $1.06$ $\pm$ $ 0.09$ $M_{\sun}$ and with an age range of 0.5 to 7 Gyrs.

\item The BSS radial distribution shows a bimodal trend with a peak 
in the centre, a minimum at $r$ $\sim$ $r_c$, and an external rising trend followed by a flattening in the outermost region of the cluster. We also determined $A^{+}_{rh}$ parameter, which is found to be $0.038$ $\pm$ $0.016$. This newly determined parameter along with the bimodal trend in the BSS radial distribution indicate that this cluster is an intermediate dynamical age cluster. Our results are consistent with the \textit{empirical dynamical clock} relation defined in F12 and L16, further confirming that NGC 6656 is an intermediate dynamical-aged cluster belonging to \textit{Family II} classification of GCs.     
\end{enumerate}
\section*{ACKNOWLEDGEMENT} 
We thank the anonymous referee for the useful comments that helped us to improve the scientific content of the paper. We would like to acknowledge ESO, for the archival data observed with ESO Telescope at the La Silla Observatory under program ID 163.O-0741(C) and 088.A-9012(A). This work has made use of data from the European Space Agency (ESA) mission {\it Gaia} (\url{https://www.cosmos.esa.int/gaia}), processed by the {\it Gaia} Data Processing and Analysis Consortium (DPAC, \url{https://www.cosmos.esa.int/web/gaia/dpac/consortium}). Funding for the DPAC has been provided by national institutions, in particular, the institutions participating in the {\it Gaia} Multilateral Agreement. We are thankful to Dr. Andrea Bellini for the description of the astro-photometric catalogue.

\bibliographystyle{mnras}
\bibliography{gaurav}

\begin{thebibliography}{}
\makeatletter
\relax
\def\mn@urlcharsother{\let\do\@makeother \do\$\do\&\do\#\do\^\do\_\do\%\do\~}
\def\mn@doi{\begingroup\mn@urlcharsother \@ifnextchar [ {\mn@doi@}
  {\mn@doi@[]}}
\def\mn@doi@[#1]#2{\def\@tempa{#1}\ifx\@tempa\@empty \href
  {http://dx.doi.org/#2} {doi:#2}\else \href {http://dx.doi.org/#2} {#1}\fi
  \endgroup}
\def\mn@eprint#1#2{\mn@eprint@#1:#2::\@nil}
\def\mn@eprint@arXiv#1{\href {http://arxiv.org/abs/#1} {{\tt arXiv:#1}}}
\def\mn@eprint@dblp#1{\href {http://dblp.uni-trier.de/rec/bibtex/#1.xml}
  {dblp:#1}}
\def\mn@eprint@#1:#2:#3:#4\@nil{\def\@tempa {#1}\def\@tempb {#2}\def\@tempc
  {#3}\ifx \@tempc \@empty \let \@tempc \@tempb \let \@tempb \@tempa \fi \ifx
  \@tempb \@empty \def\@tempb {arXiv}\fi \@ifundefined
  {mn@eprint@\@tempb}{\@tempb:\@tempc}{\expandafter \expandafter \csname
  mn@eprint@\@tempb\endcsname \expandafter{\@tempc}}}

\bibitem[\protect\citeauthoryear{{Alessandrini}, {Lanzoni}, {Ferraro},
  {Miocchi}  \& {Vesperini}}{{Alessandrini} et~al.}{2016}]{2016ApJ...833..252A}
{Alessandrini} E.,  {Lanzoni} B.,  {Ferraro} F.~R.,  {Miocchi} P.,
  {Vesperini} E.,  2016, \mn@doi [\apj] {10.3847/1538-4357/833/2/252}, \href
  {http://adsabs.harvard.edu/abs/2016ApJ...833..252A} {833, 252}

\bibitem[\protect\citeauthoryear{{Anderson}, {Bedin}, {Piotto}, {Yadav}  \&
  {Bellini}}{{Anderson} et~al.}{2006}]{2006A&A...454.1029A}
{Anderson} J.,  {Bedin} L.~R.,  {Piotto} G.,  {Yadav} R.~S.,   {Bellini} A.,
  2006, \mn@doi [\aap] {10.1051/0004-6361:20065004}, \href
  {http://adsabs.harvard.edu/abs/2006A%26A...454.1029A} {454, 1029}

\bibitem[\protect\citeauthoryear{{Arenou} et~al.,}{{Arenou}
  et~al.}{2018}]{2018A&A...616A..17A}
{Arenou} F.,  et~al., 2018, \mn@doi [\aap] {10.1051/0004-6361/201833234}, \href
  {http://adsabs.harvard.edu/abs/2018A%26A...616A..17A} {616, A17}

\bibitem[\protect\citeauthoryear{{Bailyn}}{{Bailyn}}{1995}]{1995ARA&A..33..133B}
{Bailyn} C.~D.,  1995, \mn@doi [\araa] {10.1146/annurev.aa.33.090195.001025},
  \href {http://adsabs.harvard.edu/abs/1995ARA%26A..33..133B} {33, 133}

\bibitem[\protect\citeauthoryear{{Baldwin}, {Watkins}, {van der Marel},
  {Bianchini}, {Bellini}  \& {Anderson}}{{Baldwin}
  et~al.}{2016}]{2016ApJ...827...12B}
{Baldwin} A.~T.,  {Watkins} L.~L.,  {van der Marel} R.~P.,  {Bianchini} P.,
  {Bellini} A.,   {Anderson} J.,  2016, \mn@doi [\apj]
  {10.3847/0004-637X/827/1/12}, \href
  {http://adsabs.harvard.edu/abs/2016ApJ...827...12B} {827, 12}

\bibitem[\protect\citeauthoryear{{Beccari}, {Lanzoni}, {Ferraro}, {Pulone},
  {Bellazzini}, {Fusi Pecci}, {Rood}  \& {Giallongo}}{{Beccari}
  et~al.}{2008}]{2008ApJ...679..712B}
{Beccari} G.,  {Lanzoni} B.,  {Ferraro} F.~R.,  {Pulone} L.,  {Bellazzini} M.,
  {Fusi Pecci} F.,  {Rood} R.~T.,   {Giallongo} 2008, \mn@doi [\apj]
  {10.1086/587689}, \href {http://adsabs.harvard.edu/abs/2008ApJ...679..712B}
  {679, 712}

\bibitem[\protect\citeauthoryear{{Beccari}, {Sollima}, {Ferraro}, {Lanzoni},
  {Bellazzini}, {De Marchi}, {Valls-Gabaud}  \& {Rood}}{{Beccari}
  et~al.}{2011}]{2011ApJ...737L...3B}
{Beccari} G.,  {Sollima} A.,  {Ferraro} F.~R.,  {Lanzoni} B.,  {Bellazzini} M.,
   {De Marchi} G.,  {Valls-Gabaud} D.,   {Rood} R.~T.,  2011, \mn@doi [\apjl]
  {10.1088/2041-8205/737/1/L3}, \href
  {http://adsabs.harvard.edu/abs/2011ApJ...737L...3B} {737, L3}

\bibitem[\protect\citeauthoryear{{Contreras Ramos}, {Ferraro}, {Dalessandro},
  {Lanzoni}  \& {Rood}}{{Contreras Ramos} et~al.}{2012}]{2012ApJ...748...91C}
{Contreras Ramos} R.,  {Ferraro} F.~R.,  {Dalessandro} E.,  {Lanzoni} B.,
  {Rood} R.~T.,  2012, \mn@doi [\apj] {10.1088/0004-637X/748/2/91}, \href
  {http://adsabs.harvard.edu/abs/2012ApJ...748...91C} {748, 91}

\bibitem[\protect\citeauthoryear{{Dalessandro}, {Lanzoni}, {Ferraro}, {Rood},
  {Milone}, {Piotto}  \& {Valenti}}{{Dalessandro}
  et~al.}{2008a}]{2008ApJ...677.1069D}
{Dalessandro} E.,  {Lanzoni} B.,  {Ferraro} F.~R.,  {Rood} R.~T.,  {Milone} A.,
   {Piotto} G.,   {Valenti} E.,  2008a, \mn@doi [\apj] {10.1086/529028}, \href
  {http://adsabs.harvard.edu/abs/2008ApJ...677.1069D} {677, 1069}

\bibitem[\protect\citeauthoryear{{Dalessandro}, {Lanzoni}, {Ferraro}, {Vespe},
  {Bellazzini}  \& {Rood}}{{Dalessandro} et~al.}{2008b}]{2008ApJ...681..311D}
{Dalessandro} E.,  {Lanzoni} B.,  {Ferraro} F.~R.,  {Vespe} F.,  {Bellazzini}
  M.,   {Rood} R.~T.,  2008b, \mn@doi [\apj] {10.1086/588462}, \href
  {http://adsabs.harvard.edu/abs/2008ApJ...681..311D} {681, 311}

\bibitem[\protect\citeauthoryear{{Ferraro}, {Pecci}, {Cacciari}, {Corsi},
  {Buonanno}, {Fahlman}  \& {Richer}}{{Ferraro}
  et~al.}{1993}]{1993AJ....106.2324F}
{Ferraro} F.~R.,  {Pecci} F.~F.,  {Cacciari} C.,  {Corsi} C.,  {Buonanno} R.,
  {Fahlman} G.~G.,   {Richer} H.~B.,  1993, \mn@doi [\aj] {10.1086/116804},
  \href {http://adsabs.harvard.edu/abs/1993AJ....106.2324F} {106, 2324}

\bibitem[\protect\citeauthoryear{{Ferraro}, {D'Amico}, {Possenti}, {Mignani}
  \& {Paltrinieri}}{{Ferraro} et~al.}{2001}]{2001ApJ...561..337F}
{Ferraro} F.~R.,  {D'Amico} N.,  {Possenti} A.,  {Mignani} R.~P.,
  {Paltrinieri} B.,  2001, \mn@doi [\apj] {10.1086/322773}, \href
  {http://adsabs.harvard.edu/abs/2001ApJ...561..337F} {561, 337}

\bibitem[\protect\citeauthoryear{{Ferraro}, {Beccari}, {Rood}, {Bellazzini},
  {Sills}  \& {Sabbi}}{{Ferraro} et~al.}{2004}]{2004ApJ...603..127F}
{Ferraro} F.~R.,  {Beccari} G.,  {Rood} R.~T.,  {Bellazzini} M.,  {Sills} A.,
  {Sabbi} E.,  2004, \mn@doi [\apj] {10.1086/381229}, \href
  {http://adsabs.harvard.edu/abs/2004ApJ...603..127F} {603, 127}

\bibitem[\protect\citeauthoryear{{Ferraro}, {Sollima}, {Rood}, {Origlia},
  {Pancino}  \& {Bellazzini}}{{Ferraro} et~al.}{2006}]{2006ApJ...638..433F}
{Ferraro} F.~R.,  {Sollima} A.,  {Rood} R.~T.,  {Origlia} L.,  {Pancino} E.,
  {Bellazzini} M.,  2006, \mn@doi [\apj] {10.1086/498735}, \href
  {http://adsabs.harvard.edu/abs/2006ApJ...638..433F} {638, 433}

\bibitem[\protect\citeauthoryear{{Ferraro} et~al.,}{{Ferraro}
  et~al.}{2012}]{2012Natur.492..393F}
{Ferraro} F.~R.,  et~al., 2012, \mn@doi [\nat] {10.1038/nature11686}, \href
  {http://adsabs.harvard.edu/abs/2012Natur.492..393F} {492, 393}

\bibitem[\protect\citeauthoryear{{Gaia Collaboration} et~al.,}{{Gaia
  Collaboration} et~al.}{2016}]{2016A&A...595A...1G}
{Gaia Collaboration} et~al., 2016, \mn@doi [\aap]
  {10.1051/0004-6361/201629272}, \href
  {http://adsabs.harvard.edu/abs/2016A%26A...595A...1G} {595, A1}

\bibitem[\protect\citeauthoryear{{Gaia Collaboration} et~al.,}{{Gaia
  Collaboration} et~al.}{2018}]{2018A&A...616A...1G}
{Gaia Collaboration} et~al., 2018, \mn@doi [\aap]
  {10.1051/0004-6361/201833051}, \href
  {http://adsabs.harvard.edu/abs/2018A%26A...616A...1G} {616, A1}

\bibitem[\protect\citeauthoryear{{Girardi}, {Bertelli}, {Bressan}, {Chiosi},
  {Groenewegen}, {Marigo}, {Salasnich}  \& {Weiss}}{{Girardi}
  et~al.}{2002}]{2002A&A...391..195G}
{Girardi} L.,  {Bertelli} G.,  {Bressan} A.,  {Chiosi} C.,  {Groenewegen}
  M.~A.~T.,  {Marigo} P.,  {Salasnich} B.,   {Weiss} A.,  2002, \mn@doi [\aap]
  {10.1051/0004-6361:20020612}, \href
  {http://adsabs.harvard.edu/abs/2002A%26A...391..195G} {391, 195}

\bibitem[\protect\citeauthoryear{{Goldsbury}, {Richer}, {Anderson}, {Dotter},
  {Sarajedini}  \& {Woodley}}{{Goldsbury} et~al.}{2010}]{2010AJ....140.1830G}
{Goldsbury} R.,  {Richer} H.~B.,  {Anderson} J.,  {Dotter} A.,  {Sarajedini}
  A.,   {Woodley} K.,  2010, \mn@doi [\aj] {10.1088/0004-6256/140/6/1830},
  \href {http://adsabs.harvard.edu/abs/2010AJ....140.1830G} {140, 1830}

\bibitem[\protect\citeauthoryear{{Harris}}{{Harris}}{2010}]{2010arXiv1012.3224H}
{Harris} W.~E.,  2010, preprint, \href
  {http://adsabs.harvard.edu/abs/2010arXiv1012.3224H} {} (\mn@eprint {arXiv}
  {1012.3224})

\bibitem[\protect\citeauthoryear{{Hypki} \& {Giersz}}{{Hypki} \&
  {Giersz}}{2017}]{2017MNRAS.471.2537H}
{Hypki} A.,  {Giersz} M.,  2017, \mn@doi [\mnras] {10.1093/mnras/stx1718},
  \href {http://adsabs.harvard.edu/abs/2017MNRAS.471.2537H} {471, 2537}

\bibitem[\protect\citeauthoryear{{Jordi} et~al.,}{{Jordi}
  et~al.}{2010}]{2010A&A...523A..48J}
{Jordi} C.,  et~al., 2010, \mn@doi [\aap] {10.1051/0004-6361/201015441}, \href
  {http://adsabs.harvard.edu/abs/2010A%26A...523A..48J} {523, A48}

\bibitem[\protect\citeauthoryear{{King}}{{King}}{1962}]{1962AJ.....67..471K}
{King} I.,  1962, \mn@doi [\aj] {10.1086/108756}, \href
  {http://adsabs.harvard.edu/abs/1962AJ.....67..471K} {67, 471}

\bibitem[\protect\citeauthoryear{{Kunder} et~al.,}{{Kunder}
  et~al.}{2014}]{2014A&A...572A..30K}
{Kunder} A.,  et~al., 2014, \mn@doi [\aap] {10.1051/0004-6361/201424113}, \href
  {http://adsabs.harvard.edu/abs/2014A%26A...572A..30K} {572, A30}

\bibitem[\protect\citeauthoryear{{Lanzoni}, {Dalessandro}, {Ferraro},
  {Mancini}, {Beccari}, {Rood}, {Mapelli}  \& {Sigurdsson}}{{Lanzoni}
  et~al.}{2007a}]{2007ApJ...663..267L}
{Lanzoni} B.,  {Dalessandro} E.,  {Ferraro} F.~R.,  {Mancini} C.,  {Beccari}
  G.,  {Rood} R.~T.,  {Mapelli} M.,   {Sigurdsson} S.,  2007a, \mn@doi [\apj]
  {10.1086/518592}, \href {http://adsabs.harvard.edu/abs/2007ApJ...663..267L}
  {663, 267}

\bibitem[\protect\citeauthoryear{{Lanzoni} et~al.,}{{Lanzoni}
  et~al.}{2007b}]{2007ApJ...663.1040L}
{Lanzoni} B.,  et~al., 2007b, \mn@doi [\apj] {10.1086/518688}, \href
  {http://adsabs.harvard.edu/abs/2007ApJ...663.1040L} {663, 1040}

\bibitem[\protect\citeauthoryear{{Lanzoni}, {Dalessandro}, {Perina}, {Ferraro},
  {Rood}  \& {Sollima}}{{Lanzoni} et~al.}{2007c}]{2007ApJ...670.1065L}
{Lanzoni} B.,  {Dalessandro} E.,  {Perina} S.,  {Ferraro} F.~R.,  {Rood} R.~T.,
    {Sollima} A.,  2007c, \mn@doi [\apj] {10.1086/522301}, \href
  {http://adsabs.harvard.edu/abs/2007ApJ...670.1065L} {670, 1065}

\bibitem[\protect\citeauthoryear{{Lanzoni}, {Ferraro}, {Alessandrini},
  {Dalessandro}, {Vesperini}  \& {Raso}}{{Lanzoni}
  et~al.}{2016}]{2016ApJ...833L..29L}
{Lanzoni} B.,  {Ferraro} F.~R.,  {Alessandrini} E.,  {Dalessandro} E.,
  {Vesperini} E.,   {Raso} S.,  2016, \mn@doi [\apjl]
  {10.3847/2041-8213/833/2/L29}, \href
  {http://adsabs.harvard.edu/abs/2016ApJ...833L..29L} {833, L29}

\bibitem[\protect\citeauthoryear{{Meylan} \& {Heggie}}{{Meylan} \&
  {Heggie}}{1997}]{1997A&ARv...8....1M}
{Meylan} G.,  {Heggie} D.~C.,  1997, \mn@doi [\aapr] {10.1007/s001590050008},
  \href {http://adsabs.harvard.edu/abs/1997A%26ARv...8....1M} {8, 1}

\bibitem[\protect\citeauthoryear{{Miocchi}, {Pasquato}, {Lanzoni}, {Ferraro},
  {Dalessandro}, {Vesperini}, {Alessandrini}  \& {Lee}}{{Miocchi}
  et~al.}{2015}]{2015ApJ...799...44M}
{Miocchi} P.,  {Pasquato} M.,  {Lanzoni} B.,  {Ferraro} F.~R.,  {Dalessandro}
  E.,  {Vesperini} E.,  {Alessandrini} E.,   {Lee} Y.-W.,  2015, \mn@doi [\apj]
  {10.1088/0004-637X/799/1/44}, \href
  {http://adsabs.harvard.edu/abs/2015ApJ...799...44M} {799, 44}

\bibitem[\protect\citeauthoryear{{Montegriffo}, {Ferraro}, {Fusi Pecci}  \&
  {Origlia}}{{Montegriffo} et~al.}{1995}]{1995MNRAS.276..739M}
{Montegriffo} P.,  {Ferraro} F.~R.,  {Fusi Pecci} F.,   {Origlia} L.,  1995,
  \mn@doi [\mnras] {10.1093/mnras/276.3.739}, \href
  {http://adsabs.harvard.edu/abs/1995MNRAS.276..739M} {276, 739}

\bibitem[\protect\citeauthoryear{{Paresce}, {Meylan}, {Shara}, {Baxter}  \&
  {Greenfield}}{{Paresce} et~al.}{1991}]{1991Natur.352..297P}
{Paresce} F.,  {Meylan} G.,  {Shara} M.,  {Baxter} D.,   {Greenfield} P.,
  1991, \mn@doi [\nat] {10.1038/352297a0}, \href
  {http://adsabs.harvard.edu/abs/1991Natur.352..297P} {352, 297}

\bibitem[\protect\citeauthoryear{{Piotto} et~al.,}{{Piotto}
  et~al.}{2015}]{2015AJ....149...91P}
{Piotto} G.,  et~al., 2015, \mn@doi [\aj] {10.1088/0004-6256/149/3/91}, \href
  {http://adsabs.harvard.edu/abs/2015AJ....149...91P} {149, 91}

\bibitem[\protect\citeauthoryear{{Renzini} \& {Fusi Pecci}}{{Renzini} \& {Fusi
  Pecci}}{1988}]{1988ARA&A..26..199R}
{Renzini} A.,  {Fusi Pecci} F.,  1988, \mn@doi [\araa]
  {10.1146/annurev.aa.26.090188.001215}, \href
  {http://adsabs.harvard.edu/abs/1988ARA%26A..26..199R} {26, 199}

\bibitem[\protect\citeauthoryear{{Sabbi}, {Ferraro}, {Sills}  \&
  {Rood}}{{Sabbi} et~al.}{2004}]{2004ApJ...617.1296S}
{Sabbi} E.,  {Ferraro} F.~R.,  {Sills} A.,   {Rood} R.~T.,  2004, \mn@doi
  [\apj] {10.1086/425492}, \href
  {http://adsabs.harvard.edu/abs/2004ApJ...617.1296S} {617, 1296}

\bibitem[\protect\citeauthoryear{{Sandage}}{{Sandage}}{1953}]{1953AJ.....58...61S}
{Sandage} A.~R.,  1953, \mn@doi [\aj] {10.1086/106822}, \href
  {http://adsabs.harvard.edu/abs/1953AJ.....58...61S} {58, 61}

\bibitem[\protect\citeauthoryear{{Sanna}, {Dalessandro}, {Ferraro}, {Lanzoni},
  {Miocchi}  \& {O'Connell}}{{Sanna} et~al.}{2014}]{2014ApJ...780...90S}
{Sanna} N.,  {Dalessandro} E.,  {Ferraro} F.~R.,  {Lanzoni} B.,  {Miocchi} P.,
   {O'Connell} R.~W.,  2014, \mn@doi [\apj] {10.1088/0004-637X/780/1/90}, \href
  {http://adsabs.harvard.edu/abs/2014ApJ...780...90S} {780, 90}

\bibitem[\protect\citeauthoryear{{Shara}, {Saffer}  \& {Livio}}{{Shara}
  et~al.}{1997}]{1997ApJ...489L..59S}
{Shara} M.~M.,  {Saffer} R.~A.,   {Livio} M.,  1997, \mn@doi [\apjl]
  {10.1086/310952}, \href {http://adsabs.harvard.edu/abs/1997ApJ...489L..59S}
  {489, L59}

\bibitem[\protect\citeauthoryear{{Soto} et~al.,}{{Soto}
  et~al.}{2017}]{2017AJ....153...19S}
{Soto} M.,  et~al., 2017, \mn@doi [\aj] {10.3847/1538-3881/153/1/19}, \href
  {http://adsabs.harvard.edu/abs/2017AJ....153...19S} {153, 19}

\bibitem[\protect\citeauthoryear{{Trager}, {King}  \& {Djorgovski}}{{Trager}
  et~al.}{1995}]{1995AJ....109..218T}
{Trager} S.~C.,  {King} I.~R.,   {Djorgovski} S.,  1995, \mn@doi [\aj]
  {10.1086/117268}, \href {http://adsabs.harvard.edu/abs/1995AJ....109..218T}
  {109, 218}

\makeatother
\end{thebibliography}

\label{lastpage}

\end {document}